\documentclass[twocolumn, pra , nofootinbib , superscriptaddress
]{revtex4-1}
\usepackage{graphicx}
\usepackage{amsfonts}
\usepackage{amssymb}
\usepackage{amsthm}
\usepackage{array}
\usepackage{float}
\usepackage{amsmath}
\usepackage{verbatim} 
\usepackage{hyperref}
\usepackage{color}
\usepackage{bbold}
\usepackage{epstopdf}
\usepackage{mathtools}
\usepackage{enumerate}
\usepackage{subcaption}
\usepackage{ulem}
\usepackage{physics}
\usepackage{subcaption}
\usepackage{soul,xcolor}
\usepackage{newtxtext,newtxmath}
\usepackage{hyphenat}

\newtheorem*{proposition*}{Proposition}

\newtheorem{theorem}{Theorem}
\newtheorem*{theorem*}{Theorem}

\newtheorem*{corollary*}{Corollary}

\def\Tr{\mathrm{Tr}}
\setstcolor{red}

\begin{document}
\title{Quantum stochastic series expansion methods}
\author{Kok Chuan Tan}
\email{bbtankc@gmail.com}
\affiliation{ School of Physical and Mathematical Sciences, Nanyang Technological University, Singapore 637371, Republic of Singapore}
\author{Dhiman Bowmick}
\affiliation{ School of Physical and Mathematical Sciences, Nanyang Technological University, Singapore 637371, Republic of Singapore}
\author{Pinaki Sengupta}
\affiliation{ School of Physical and Mathematical Sciences, Nanyang Technological University, Singapore 637371, Republic of Singapore}

\begin{abstract}
A quantum implementation of the Stochastic Series Expansion (SSE) Monte Carlo method is proposed, and it is shown that quantum SSE offers significant advantages over classical implementations of SSE. In particular, for problems where classical SSE encounters the sign problem, the cost of implementing a Monte Carlo iteration scales only linearly with system size in quantum SSE, while it may scale exponentially with system size in classical SSE. In cases where classical SSE can be efficiently implemented, quantum SSE still offers an advantage by allowing for more general observables to be measured.
\end{abstract}

\maketitle

\section{Introduction}

The Stochastic Series Expansion (SSE)\cite{Sandvik1991,Sandvik1992, Sandvik1997, Sandvik1999} method is a widely used Quantum Monte Carlo (QMC) method for simulating models of quantum many-body systems. It is based on sampling the series expansion of $\exp(-\beta H)$ up to a sufficiently high order. A significant advantage of SSE is that expectation values that are obtained via this method are exact, up to statistical errors. Alternative approaches include the world line method \cite{Hirsch1982,Suzuki1976, Suzuki1977, Beard1996}, and the DMRG method \cite{Beard1996, White1992}. In this article, we compare implementations of SSE method on a quantum computer to its implementation on a classical computer. The former is here referred to as quantum SSE and the latter as classical SSE. Quantum computers are promising platforms to speed up the simulation of quantum many-body systems. Algorithms that exploit quantum hardware to speed up simulations of the thermal Gibbs state of many-body systems have previously been explored in Refs.~\cite{Terhal2000,Bilgin2010, Temme2011, Riera2012, Yung2012, Montanaro2015, Ge2016, Motta2019, Tan2020}. We will demonstrate several advantages that quantum SSE has over classical SSE. In particular, we argue that the ``no-branching" requirement\cite{Sandvik2010} of classical SSE can be relaxed in quantum SSE, which leads to important consequences for the simulation of many-body systems. 

First,  lifting the ``no-branching" requirement in quantum SSE allows for the use of arbitrary superpositions of states. This means that we are no longer limited to basis states that permit a diagonal representation. This has the effect of allowing more general quantum observables to be measured in quantum SSE. 

The second consequence is that quantum SSE always leads to nonnegative weights, which are directly sampled via measurement probabilities. Furthermore, these nonnegative weights can be sampled in polynomial time. This implies that quantum computers may be able to simulate many-body systems currently inaccessible to classical SSE methods due to the famous ``sign problem"\cite{Foulkes2001, Henelius2000}. Notably, the Quantum Metropolis Sampling (QMS)\cite{Temme2011} algorithm also avoids the sign problem by repeated use of the quantum phase estimation algorithm\cite{Abrams1999}. However, quantum phase estimation requires deep quantum circuits, and approximates the unitary operation $U = \exp(iHt)$ via the Suzuki-Trotter decomposition\cite{Lloyd1996}. This necessarily introduces a systematic error, unlike exact QMC methods such as SSE, which does not involve Trotterization.

This article is structured as follows: First, we introduce the broad ideas underlying the SSE QMC method. Second, we will describe a possible SSE implementation on a quantum computer, first for a simpler special case, then for the more general case. Third, we discuss how the sign problem affects classical SSE. Fourth, we summarize and evaluate the advantages that quantum SSE offers over classical SSE. Finally, we numerically simulate the quantum SSE algorithm for one dimensional spin chains and compare it with exact results.

\section{Preliminaries}
We briefly review the Stochastic Series Expansion (SSE) method\cite{Sandvik2010}. Let us consider a system with Hamiltonian $H'$ in thermal thermal equilibrium at inverse temperature $\beta$. The partition function is defined as
\begin{align}
Z \coloneqq \Tr(e^{-\beta H'}) = \sum_{\alpha} \bra{\alpha}e^{-\beta H'}\ket{\alpha},
\end{align} where $\{ \ket{\alpha} \}$ is some complete set of basis vectors.

We are interested to find the value of some observable $O$ for this system. In general, this is given by 
\begin{align}
\expval{O} = \Tr(O e^{-\beta H'})/Z.
\end{align} Generally speaking, the matrix $e^{-\beta H'}$ is difficult to compute, especially for large system sizes. One way to make the problem more tractable is to perform a series expansion of the matrix exponent. Let $H \coloneqq -H'$. We have:
\begin{align}
e^{\beta H} = \sum_{n=0}^\infty \frac{\beta^n}{n!} H^n.
\label{eq::Boltzmann}
\end{align} 
Depending on the system, it may be convenient to further decompose the Hamiltonian such that $H = \sum_{b} H_b$. This allows us to write 
\begin{align}
e^{\beta H} &= \sum_{n=0}^\infty \frac{\beta^n}{n!} (\sum_{b_n} H_{b_n}) \ldots (\sum_{b_1} H_{b_1}) \\
&= \sum_{n=0}^\infty \sum_{b} \frac{\beta^n}{n!} H_{b_n} \ldots H_{b_1},
\end{align} where $b$ denotes the operator string $b_n \ldots b_1$. The partition function can then be written as
\begin{align}
Z= \sum_{n=0}^\infty \sum_{b} \sum_{\alpha} \frac{\beta^n}{n!} \bra{\alpha}H_{b_n} \ldots H_{b_1} \ket{\alpha}.
\end{align}

Assuming that each term $\bra{\alpha}H_{b_n} \ldots H_{b_1} \ket{\alpha}$ is nonnegative, the idea behind SSE is to perform a Quantum Monte Carlo (QMC) simulation by randomly sampling the configuration space $\mathcal{C} \coloneqq \{ (n, b, \alpha)   \; \forall \; n, b, \alpha \} $. The effective SSE partition function being sampled is given by
\begin{align}
Z_{\mathrm{SSE}} &= \sum_{n=0}^M \sum_{b}\sum_{\alpha}  \frac{\beta^n}{n!}\bra{\alpha}H_{b_n}\ldots H_{b_1}\ket{\alpha}  \label{eq::SSEPartitionFun}\\
&= \sum_{C} \frac{\beta^n}{n!}\bra{\alpha}H_{b_n}\ldots H_{b_1}\ket{\alpha},
\end{align} where $M$ is some sufficiently large cutoff in the expansion power. In order to obtain the expectation value of some observable $O$, we need to find some function $f(O,C)$ which gives an unbiased estimate of $\expval{O}$ such that 
\begin{align}
\expval{f(O,C)} = \sum_C p_C f(O,C) = \expval{O}, \label{eq::expVal}
\end{align} where $p_C \coloneqq \frac{\beta^n}{n!}\bra{\alpha}H_{b_n}\ldots H_{b_1}\ket{\alpha} / Z_{\text{SSE}}$. Finding $\expval{f(O,C)}$ for a given $O$ is not necessarily trivial, but for the case where $O$ is a diagonal matrix, we see that 
\begin{align}
\expval{O} &= \Tr(O e^{\beta H})/Z_{\text{SSE}} \\
&= \sum_\alpha \bra{\alpha }O \ket{\alpha}  \bra {\alpha }e^{\beta H} \ket{\alpha}/Z_{\text{SSE}} \\
&= \sum_C  p_C \bra{\alpha }O \ket{\alpha}, 
\end{align} so $f(O,C) \coloneqq \bra{\alpha }O \ket{\alpha} $ is an example of an unbiased estimator.

\section{SSE on a quantum computer} \label{sec::quantumSSE}

We now propose a method of implementing a SSE Monte Carlo simulation on a quantum computer. 

Let us consider a decomposition of the Hamiltonian $H = \sum_b H_b$. The classical implementation of the SSE method requires that $H_b$ satisfy a  so-called "no-branching" condition in order for the algorithm to be efficient  (see Section~\ref{sec::classSSE}). On a quantum computer however, this requirement is no longer necessary as quantum computers naturally allows for superpositions of a large number of states. We can therefore choose a more convenient decomposition. In general, it is always possible to decompose any Hamiltonian as a sum of products of Pauli matrices: 
\begin{align}
H = \sum_b h_b \bigotimes_{i=1}^N \sigma^{(i)}_{b^i}, \label{eq::hamDecomp}
\end{align} where in general $b^i = 0,1,2,3$ and $\sigma_0 \coloneqq \openone $, $\sigma_1 \coloneqq \sigma_x $, $\sigma_2 \coloneqq \sigma_y $ and $\sigma_3 \coloneqq \sigma_z $. Note that in this notation, we used the upper index to label the Pauli matrices. This is different from the lower index used to label the operator string $b$ in $H_{b_n} \ldots H_{b_1}$.

In order to illustrate the quantum SSE method, we first consider a  special case where the operators $h_{b_i}$ mutually commutes. For example, this can occur when we restrict ourselves to $b^i = 0,1$, such that the only Pauli matrices we need to consider are $\openone$ and $\sigma_x$. Such problems can already be nontrivial. For instance, in Ref.~\cite{Troyer2005}, the Hamiltonian
\begin{align} \label{eq::XXInter}
H' = -\sum_{\expval{j,k}} J_{jk}\sigma_x^{(j)} \sigma_x^{(k)},
\end{align} was considered as an example of a many-body system that is NP hard to simulate for certain lattice configurations.

In classical SSE, the basis $\{ \ket{\alpha} \}$ must be chosen carefully so that the Monte Carlo sampling is efficient. In this section, we describe how the quantum implementation of SSE can lift this restriction, and allow for the use of more general $\{ \ket{\alpha} \}$.

We first define $H_b =  h_b \bigotimes_{i=1}^N \sigma^{(i)}_{b^i} + \abs{h_b} \openone$, which ensures that $H_b$ is always positive semidefinite. We can verify that 
\begin{align}
&\frac{H_b H _{b'}}{\abs{h_b h_{b'}}} \\
&= (\openone + \text{sgn}(h_b)\bigotimes_{i=1}^N \sigma^{(i)}_{b_i} )(\openone + \text{sgn}(h_{b'})\bigotimes_{i=1}^N  \sigma^{(i)}_{b'_i} ) \\
&=\openone + \text{sgn}(h_b)\bigotimes_{i=1}^N \sigma^{(i)}_{b_i} + \text{sgn}(h_{b'})\bigotimes_{i=1}^N  \sigma^{(i)}_{b'_i} \notag  \\
&\quad +  \text{sgn}(h_{b}) \text{sgn}(h_{b'})\bigotimes_{i=1}^N  \sigma^{(i)}_{b_i}\sigma^{(i)}_{b'_i} \\
&= (\openone + \text{sgn}(h_{b'})\bigotimes_{i=1}^N  \sigma^{(i)}_{b'_i} ) (\openone + \text{sgn}(h_b)\bigotimes_{i=1}^N \sigma^{(i)}_{b_i} )\\
&= \frac{H_{b'} H _{b}}{\abs{h_{b'} h_{b}}},
\end{align} where we used the fact that $\sigma_{b^i}^{(i)}$ can only be either be $\openone$ or $\sigma_x$ and they mutually commute. Therefore, $H_b$ forms a set of mutually commuting observables. 

This implies that a product of such operators $H_{b_n} \ldots H_{b_1}$ is also positive semidefinite. To see this, suppose $A$ and $B$ are positive Hermitian operators that commute. This means $AB$ is Hermitian since $(AB)^\dag = B^\dag A^\dag = AB$. We then observe that $AB$ must have the same eigenvalues as $A^{1/2} B A^{1/2}$. This is because if $A^{1/2} B A^{1/2} \ket{\lambda} = \lambda \ket{\lambda}$, then $ A^{1/2}\ket{\lambda}$ must be an eigenvector of $AB$ with the same eigenvalue, since $ AB(A^{1/2}\ket{\lambda}) = A^{1/2}(A^{1/2}BA^{1/2}) \ket{\lambda} = \lambda A^{1/2} \ket{\lambda}$. Since $A^{1/2}BA^{1/2}$ is positive semidefinite, $AB$ must therefore be positive semidefinite.

Making $H_b$ positive semidefinite is equivalent to adding a constant to the Hamiltonian
\begin{align}
H \rightarrow H + k \openone, \label{eq::positiveHam}
\end{align} where $k \coloneqq \sum_b \abs{h_b}$, such that the total Hamiltonian is also positive semidefinite. With the positivity of $\bra{\alpha}H_{b_n} \ldots H_{b_1} \ket{\alpha}$ assured, we need a method of sampling the relative weight of a given configuration $(n,b,\alpha)$.

Let us consider a state of (N+n) qubits of the form:
\begin{align}
\ket{\alpha_{A}}  \ket{+_{B_1}} \ldots \ket{+_{B_n}},
\end{align} where $\ket{+} \coloneqq (\ket{0}+\ket{1})/\sqrt{2}$, $N$ is the number of particles in the system we are trying to simulate, $n$ is the expansion power in the SSE, and $A = A_1 \ldots A_N$. 

Observe that $H_b =  \abs{h_b} \left [\text{sgn}(h_b)\bigotimes_{i=1}^N \sigma^{(A_i)}_{b^i} + \openone_A \right ]$ is a superposition of 2 unitary operators $\text{sgn}(h_b)\bigotimes_{i=1}^N \sigma^{(A_i)}_{b^i}$ and $\openone_A$. We define the following controlled unitary operation:
\begin{align}
&U_{A,B_i} \ket{\alpha_{A}} \ket{0_{B_i}} = \openone_A \ket{\alpha_A} \ket{0_{B_i}}  \label{eq::unitary1}\\
&U_{A,B_i} \ket{\alpha_A} \ket{1_{B_i}} = \left [ \text{sgn}(h_b)\bigotimes_{j=1}^N \sigma^{(A_i)}_{b^j} \right ]\ket{\alpha_{A}} \ket{1_{B_i}}. \label{eq::unitary2}
\end{align} 

For illustrative purposes, consider the case where the expansion power is $n=1$. Applying $U_{A,B_1}$, we get:
\begin{align}
U_{A,B_1} & \ket{\alpha_{A}} \ket{+_{B_1}} \\   &= \frac{1}{\sqrt{2}} \ket{\alpha_A} \ket{0_{B_1}} + \frac{1}{\sqrt{2}} \text{sgn}(h_b)\bigotimes_{j=1}^N \sigma^{(A_i)}_{b^j} \ket{\alpha_{A}} \ket{1_{B_1}}.
\end{align} Projecting this onto $\ket{\alpha_A}\ket{+_{B_1}}$, we obtain the amplitude
\begin{align}
&\bra{\alpha_A} \bra{+_{B_1}}U_{A,B_1}  \ket{\alpha_{A}} \ket{+_{B_1}}\\
&= \frac{1}{2} \bra{\alpha_A} \openone_A \ket{\alpha_A} + \frac{1}{2}  \bra{\alpha_{A}}\text{sgn}(h_b)\bigotimes_{j=1}^N \sigma^{(A_i)}_{b^j} \ket{\alpha_{A}} \\
&= \frac{1}{2 \abs{h_b}} \bra{\alpha_A} H_{b} \ket{\alpha_{A}}.
\end{align}

Similarly, for arbitrary expansion powers $n$, we get
\begin{align}
&\bra{\alpha_A} \bra{+_{B_1}} \ldots \bra{+_{B_n}} U_{A,B_n}  \ldots U_{A,B_1}  \ket{\alpha_{A}} \ket{+_{B_1}} \ldots \ket{+_{B_n}}\nonumber\\
&= \frac{1}{2^n \abs{h_{b_n} \ldots h_{b_1}}} \bra{\alpha_A}H_{b_n} \ldots H_{b_1} \ket{\alpha_A}.
\label{eq::expectation}
\end{align} Note that the spectrum of $H_{b_i}/\abs{h_{b_i}}$  is in the range $[0,2]$ so the spectrum of $H_{b_n} \ldots H_{b_1} / \abs{h_{b_n} \ldots h_{b_1}}$ is within $[0,2^n]$. The projected amplitude is therefore not necessarily exponentially small even for relatively large expansion orders $n$, despite the $1/2^n$ factor. We shall also see that this factor cancels out during the Metropolis portion of the Monte Carlo simulation, where only the ratio between the configuration weights, and not the actual weight itself, matters.

At this juncture, one just needs to sample the probability 
\begin{align} 
q(n,b,\alpha) \coloneqq \abs{\frac{\bra{\alpha_A}H_{b_n} \ldots H_{b_1} \ket{\alpha_A}}{2^n h_{b_n} \ldots h_{b_1}}}^2 . \label{eq::sampProb}
\end{align} For $t$ independent samples, the sample variance scales with $ \sim 1/t$.  In this way, the configuration weights can be estimated to any target degree of numerical precision. 

Alternatively, we can also perform a quantum subroutine called amplitude estimation\cite{Brassard2002} (see Appendix) to obtain the required amplitude to any degree of precision. In general, to estimate the probability $p$ to any desired precision $\epsilon$ with success probability $1-\delta$, the subroutine needs to be invoke certain unitary operations a total of $t= t(\epsilon, \delta)$ times, where $t(\epsilon, \delta)$ only depends on the desired precision $\epsilon$ and success probability $1-\delta$. In this case, the variance scales with $\sim 1/t^2$, where $t$ is now the number of times the unitary operations are applied rather than the number of independent samples.

\section{Applying the Metropolis method}  
\label{sec::metropolis}
Once the relative weight a some configuration $C$ is sampled, the Monte Carlo simulation proceeds by implementing the Metropolis method. This consists of randomly selecting some new configuration $C'$, and then accepting the newly chosen configuration with probability
\begin{align}
P_{\text{accept}}(C \rightarrow C') \coloneqq \min \left ( \frac{W(C')}{W(C)}, 1\right ), \label{eq::metropolis}
\end{align} where $W(C)$ is the relative weight assigned to a configuration $C = (n,b,\alpha)$. It is given by the following expression
\begin{align}
W(C) &= W(n,b, \alpha) \\
&\coloneqq \frac{\beta^n}{n!}\bra{\alpha}H_{b_n}\ldots H_{b_1}\ket{\alpha} \\
&= \frac{\beta^n}{n!} \abs{2^n h_{b_n} \ldots h_{b_1}} \sqrt{q(n,b,\alpha)},
\label{eq::weight}
\end{align} where $q(n,b,\alpha)$ is the probability sampled in Eq.~\ref{eq::sampProb}. In Eq.~\ref{eq::metropolis}, it is implicitly assumed that the probability of selecting $C'$ when the current configuration is $C$ is the same as the probability of selecting $C$ when the current configuration is $C'$, i.e. $P_{\text{select}}(C \rightarrow C') = P_{\text{select}}(C' \rightarrow C)$.

Suppose we update the independent variables $n,b,\alpha$ separately. When updating the basis state $\alpha$, we have 
\begin{align}
W(n,b, \alpha') / W(n, b ,\alpha) = \sqrt{\frac{q(n,b,\alpha')}{q(n,b,\alpha)}}.
\end{align}

When updating the operator string $b$, the acceptance probability depends only on the operator strings $b,b'$:
\begin{align}
&W(n,b', \alpha) / W(n, b ,\alpha) \\
&= \abs{\frac{h_{b'_n} \ldots h_{b'_1}}{h_{b_n} \ldots h_{b_1}  } } \sqrt{\frac{q(n,b',\alpha)}{q(n,b,\alpha)}}.
\end{align}

Finally, we can update the expansion power by randomly choosing to increase or decrease the expansion power by one. In this case, we increase the length of the operator string by appending a randomly chosen element $b_{n+1}$ to the end, and the probability of accepting an increase depends on
\begin{align}
&W(n+1,b_{n+1}b, \alpha) / W(n, b,\alpha) \\
&= \frac{2\beta \abs{h_{b_{n+1}}}}{n+1} \sqrt{\frac{q(n+1,b_{n+1}b,\alpha)}{q(n,b,\alpha)}}.
\end{align} If we choose to decrease the expansion power, we remove the last element $b_n$ from the operator string, and the probability of accepting this decrease is
\begin{align}
&W(n-1,b, \alpha) / W(n, b,\alpha) \\
&= \frac{n}{2\beta\abs{h_{b_{n}}}} \sqrt{\frac{q(n-1,b_{n-1}\ldots b_1,\alpha)}{q(n,b,\alpha)}}.
\end{align}

From the above, we see that the acceptance probability depends on the ratio $\sqrt{\frac{q(n',b',\alpha')}{q(n,b,\alpha)}}$ in general.

\section{Quantum implementation of SSE for general Hamiltonians} \label{sec::genHam}

We have previously considered an implementation of quantum SSE for the special case where the quantity $\bra{\alpha}H_{b_n} \ldots H_{b_1} \ket{\alpha}$ is guaranteed to be nonnegative. For general Hamiltonians, this may not always be possible because the operator $H_{b_n} \ldots H_{b_1}$  is not Hermitian in general, so it does not always output a real number. In this section, we show how to overcome this obstacle. 

Recall the expression for the expectation value in Eq.~\ref{eq::expVal}, which is given by:
\begin{align}
\expval{O} &= \sum_C p_C f(O,C) \\
&= \sum_{n,b,\alpha}\frac{\beta^n}{n!}\bra{\alpha}H_{b_n}\ldots H_{b_1}\ket{\alpha} \bra{\alpha}O\ket{\alpha}/Z_{\text{SSE}}.
\end{align} We observe that the summation over all possible strings $b$ contain $\bra{\alpha}H_{b_n}\ldots H_{b_1}\ket{\alpha}$, as well as its complex conjugate  $\bra{\alpha}H_{b_1}\ldots H_{b_n}\ket{\alpha}$. Since  $\bra{\alpha}H_{b_n}\ldots H_{b_1}\ket{\alpha}+ \bra{\alpha}H_{b_1}\ldots H_{b_n}\ket{\alpha} = 2\Re{\bra{\alpha}H_{b_n}\ldots H_{b_1}\ket{\alpha}}$, we see that only the real portion of each term contributes to the expectation value. This means that we can equivalently write
\begin{align}
\expval{O} = \sum_{n,b,\alpha}\frac{\beta^n}{n!}\Re{\bra{\alpha}H_{b_n}\ldots H_{b_1}\ket{\alpha} }\bra{\alpha}O\ket{\alpha}/Z_{\text{SSE}}.
\end{align} Therefore, in order to implement quantum SSE, we only need to sample the real portion of $\bra{\alpha}H_{b_n}\ldots H_{b_1}\ket{\alpha}$ and ensure that it is nonnegative. We now show that this can be done by adding a sufficiently large constant to the Hamiltonian.

Suppose $M\geq n$ is the cutoff in the expansion power (see Eq.~\ref{eq::SSEPartitionFun}). For a fixed $M$, let $H_b \coloneqq  \abs{h_b} \left [\text{sgn}(h_b)\bigotimes_{i=1}^N \sigma^{(A_i)}_{b^i} + 2M \openone_A \right ]$. We note that this is an unequal superposition of 2 unitary operations that depends on the cutoff value $M$.

We introduce the state
\begin{align}
\ket{\psi_{\text{in}}} \coloneqq \ket{\alpha_{A}}  \ket{\phi_{B_1}} \ldots \ket{\phi_{B_n}} \ket{+_C},
\end{align} where 
\begin{align}
\ket{\phi_{B_i}} \coloneqq \sqrt{(2M)/(2M+1)} \ket{0_{B_i}} + \sqrt{1/(2M+1)} \ket{1_{B_i}},
\end{align} and $\ket{+_C} \coloneqq \frac{1}{\sqrt{2}}(\ket{0_C}+\ket{1_C})$.

As before, we define the following controlled unitary operation:
\begin{align}
&U_{A,B_i} \ket{\alpha_{A}} \ket{0_{B_i}} \coloneqq \openone_A \ket{\alpha_A} \ket{0_{B_i}}  \label{eq::unitary1}\\
&U_{A,B_i} \ket{\alpha_A} \ket{1_{B_i}} \coloneqq \left [ \text{sgn}(h_b)\bigotimes_{j=1}^N \sigma^{(A_i)}_{b^j} \right ]\ket{\alpha_{A}} \ket{1_{B_i}}. \label{eq::unitary2}
\end{align} 

Based on this, we further define the unitary $V_{AB,C}$, which is controlled by qubit $C$:
\begin{align}
&V_{AB,C} \ket{\alpha_{A}}  \ket{\phi_{B_1}} \ldots \ket{\phi_{B_n}} \ket{0_C} \notag \\
&\quad \coloneqq  U_{A,B_1}  \ldots U_{A,B_n} \ket{\alpha_{A}}  \ket{\phi_{B_1}} \ldots \ket{\phi_{B_n}} \ket{0_C}\\
&V_{AB,C} \ket{\alpha_{A}}  \ket{\phi_{B_1}} \ldots \ket{\phi_{B_n}} \ket{1_C} \notag\\
&\quad \coloneqq  U_{A,B_n}  \ldots U_{A,B_1} \ket{\alpha_{A}}  \ket{\phi_{B_1}} \ldots \ket{\phi_{B_n}} \ket{1_C}
\end{align}

For any given expansion power $n$, we can verify the expression:
\begin{align}
&\bra{\psi_{\text{in}}} V_{AB,C} \ket{\psi_{\text{in}}}\\
&= \frac{\bra{\alpha_A}H_{b_1} \ldots H_{b_n} \ket{\alpha_A}+\bra{\alpha_A}H_{b_n} \ldots H_{b_1} \ket{\alpha_A} }{2(2M+1)^n \abs{h_{b_n} \ldots h_{b_1}}} \\
&= \frac{\Re{\bra{\alpha_A}H_{b_n} \ldots H_{b_1} \ket{\alpha_A}} }{(2M+1)^n \abs{h_{b_n} \ldots h_{b_1}}}.
\end{align} 
Note that the spectrum of $H_{b_i}/\abs{h_{b_i}}$  is in the range $[0,2M+1]$ so the absolute value of $\Re{\bra{\alpha_A}H_{b_n} \ldots H_{b_1} \ket{\alpha_A}}/\abs{h_{b_n} \ldots h_{b_1}}$ is within the range $[0,(2M+1)^n]$. We see that the amplitude $\bra{\psi_{\text{in}}} V_{AB,C} \ket{\psi_{\text{in}}}$ gives us the required relative weight of the configuration.

We need to ensure that every configuration weight, and hence $\Re{\bra{\alpha_A}H_{b_n} \ldots H_{b_1} \ket{\alpha_A}}$ is always nonnegative. This is shown in the following series of inequalities:

\begin{align}
&(H_{b_1} \ldots H_{b_n}  + H_{b_n} \ldots H_{b_1} ) / \abs{h_{b_n} \ldots h_{b_1}} \\
&= \left (\text{sgn}(h_{b_1})\bigotimes_{i=1}^N \sigma^{(A_i)}_{b_{1}^i} + 2M \openone_A \right ) \times \ldots  \notag \\
&\qquad \times \left (\text{sgn}(h_{b_n})\bigotimes_{i=1}^N \sigma^{(A_i)}_{b_{n}^i} + 2M \openone_A \right ) \notag \\
&+\left (\text{sgn}(h_{b_n})\bigotimes_{i=1}^N \sigma^{(A_i)}_{b_{n}^i} + 2M \openone_A \right ) \times \ldots  \notag \\
&\qquad \times \left (\text{sgn}(h_{b_1})\bigotimes_{i=1}^N \sigma^{(A_i)}_{b_{1}^i} + 2M \openone_A \right ) \\
&= (2M)^{n} \openone + (2M)^{n-1} \text{sgn}(h_{b_1})\bigotimes_{i=1}^N \sigma^{(A_i)}_{b_{1}^i} + \ldots \\
& \quad +(2M)^{n} \openone + (2M)^{n-1} \text{sgn}(h_{b_n})\bigotimes_{i=1}^N \sigma^{(A_i)}_{b_{n}^i} + \ldots \\
&= 2[ (2 M)^n \openone -  (2M)^{n-1} A_1 - (2M)^{n-2} A_2- \ldots ] \\
&\geq 2[ (2 M)^n \openone -  (2M)^{n-1}\binom{n}{1} \openone - (2M)^{n-2}\binom{n}{2} \openone- \ldots ] \label{eq::genSSE1}\\
&\geq 2[ (2M)^n-(2M)^n /2 - (2M)^n/2^2 - \ldots] \label{eq::genSSE2} \\
&\geq 2[(2M)^n - 2(2M)^n /2] \label{eq::genSSE3} \\
&=0.
\end{align} Here, the matrices $A_k$ are Hermitian matrices that collects all the matrices that are products of $(n-k)$ identity matrices, and $k$ non-identity matrices. The products of Pauli matrices has eigenvalues whose absolute values are equal to 1, and $A_k$ is a sum of $\binom{n}{k}$ such products, so we have that $A_k \leq  \binom{n}{k} \openone$, which we used in Eq.~\ref{eq::genSSE1}. In Eq.~\ref{eq::genSSE2}, we used the property that $\binom{n}{k} \leq n^k \leq M^k$. Finally, in Eq.~\ref{eq::genSSE3}, we used the expression for the infinite sum of the geometric series, $\sum_{k=0}^\infty 1/2^k = 2$.

From the above arguments, we see that the configuration weight can be directly sampled by measuring the probability $q(n,b,\alpha) \coloneqq \abs{\bra{\psi_{\text{in}}} V_{AB,C} \ket{\psi_{\text{in}}}}^2$. The Metropolis portion of the simulation then proceeds as before, where the acceptance probability depends on the ratio $\sqrt{q(n',b',\alpha') / q(n,b,\alpha)}$. Note that the above proof finds a sufficiently large constant to add to the Hamiltonian to avoid negative weights. This constant is likely too large for many specific problems. We expect that the minimum constant that is required can be optimized on a case by case basis.

\section{Sign problem in classical SSE} \label{sec::classSSE}

We recall that implementing SSE Monte Carlo requires each term $\bra{\alpha}H_{b_n} \ldots H_{b_1} \ket{\alpha}$ in the expansion to be nonnegative. In general, this cannot be always guaranteed except for special cases. This is known as the \textit{sign problem}\cite{Foulkes2001, Henelius2000}.

For a typical classical implementation of SSE, there is a so-called ``no-branching" condition. This is the requirement that $H_B\ket{\alpha} \propto \ket{\alpha'} $, where $\ket{\alpha'}$ is again a basis vector. In other words, we always have to use a decomposition of $H = \sum_b H_b$ such that $H_b$ does not create superpositions of basis states. For any given basis, every $H_b$ satisfying the ``no-branching" requirement can be classified as a diagonal update satisfying $H_B\ket{\alpha} \propto \ket{\alpha} $ for every $\alpha$, or an off-diagonal update satisfying $H_B\ket{\alpha} \propto \ket{\alpha'} $ where $\alpha \neq \alpha'$ for some $\alpha$.

A diagonal update can always be made positive by adding a sufficiently large constant. This is because if $H_b$ is a diagonal update, then $H'_b \ket{\alpha} \coloneqq (H_b + k \openone) \ket{\alpha} \propto \ket{\alpha}$ is also a diagonal update.

On the other hand, we see that if $H_b$ is an off-diagonal update, adding a constant will necessarily create a superposition of basis states, since $(H_b + k \openone) \ket{\alpha} \propto h_{b,\alpha} \ket{\alpha'} + k \ket{\alpha}$ where $\alpha \neq \alpha'$. This means that we cannot guarantee that $H_b$ is always positive semidefinite for off-diagonal updates. This in turn implies that $\bra{\alpha}H_{b_n} \ldots H_{b_1} \ket{\alpha}$ is not necessarily positive, which is the sign problem.

From the above, we see that the sign problem exists because of the ``no-branching" requirement. If we avoid the sign problem by lifting no-branching requirement, one will have to keep track of all the off-diagonal elements of $H_{b_n} \ldots H_{b_1} \ket{\alpha}$. In the worst case, the computational resources required to keep track of an arbitrary superposition of basis states is  of the order $\order{\exp(N)}$, where $N$ is the number of particles. 

The typical way of circumventing the sign problem classically is to sample the absolute values of the probabilities and then correct for the sign. Suppose there are some configurations $C$ that $p_C$ can be negative. The typical approach is to write
\begin{align}
\expval{O} =  \frac{ \sum_C f(O,C) \text{sgn}(p_C) \abs{p_C} /\sum_{C'} \abs{p_{C'}} }{ \sum_{C} \text{sgn}(p_{C}) \abs{p_{C}} /\sum_{C'}\abs{p_{C'}} }.  
\end{align} One then sees that by keeping track of $\text{sgn}(p_C)$ the numerator and denominator can each be sampled with relative weights $\abs{p_C}$ using standard Monte Carlo techniques. If we consider $f'(O,C) \coloneqq f(O,C)\text{sgn}(p_C)$ to be the estimator then we can write
\begin{align}
\expval{O} = \sum_C f'(O,C) \frac{\abs{p_C} / S_{\text{corr}}}{  \sum_{C'}\abs{p_{C'}} }.
\end{align} That is, the positive relative weight $\abs{p_C}$ requires an additional corrective factor $S_{\text{corr}} \coloneqq  \sum_{C} \text{sgn}(p_{C}) \abs{p_{C}} /\sum_{C'}\abs{p_{C'}}$ to get the ``correct" weight. Unfortunately, the sampling uncertainty in  $1/S_{\text{corr}}$ grows exponentially with system size, so the amount of resources required to estimate the corrected weight scales exponentially \cite{Troyer2005}. This is similar to the conclusion that was reached by lifting the no-branching requirement in classical SSE.

\section{Comparison between quantum and classical SSE methods}

The primary benefit of the quantum SSE method is that it does not require the no-branching condition, as quantum computers naturally allows for the creation of superpositions of quantum states. This allows us to sample the relative weights of a given configuration directly, without needing to keep track of all the off-diagonal elements. By lifting the no-branching requirement, we can always ensure that the relative weights are nonnegative, thus also avoiding the sign problem. We have shown this for the special case where the Hamiltonian can be decomposed into products of $\openone$ or $\sigma_x$, as well as for more general Hamiltonians.

Similar to the classical SSE algorithm, the quantum SSE implementation computes statistical averages most easily when the observable $O$ is diagonal in the basis $\ket{\alpha}$. Unlike classical SSE approaches however, we are not required to impose strong assumptions on the basis states $\ket{\alpha}$ for a particular implementation of quantum SSE. For any given operator $O$, we can always choose the basis $\{ \ket{\alpha} \}$ to be the one that diagonalizes $O$, and the estimator is given by $f(O,C) = \ev{O}{\alpha}$. The only limitation is that the preparation of a state $\ket{\alpha}$ should be efficient on a quantum computer, i.e. the state can be prepared in polynomial time. Therefore, one important advantage of the quantum SSE method is that it allows for more general quantum observables to be measured. An example of this is when $O = \ketbra{\phi}$ for some known quantum state $\ket{\phi}$. In this case, $O$ is the projector onto the state $\ket{\phi}$ and $\expval{O} = \bra{\phi} e^{-\beta H'}/Z \ket{\phi}$ is the overlap between $\ket{\phi}$ and the thermal state $ e^{-\beta H'}/Z$. In general, finding the state overlap is not easily implementable using classical SSE. In the Shastry-Sutherland model\cite{Shastry1981, Richter1998, Miyahara2003} for instance, this can be used to directly verify that the ground state is a product of singlet pairs. This is achieved by by letting $\ket{\phi}$ be a product of singlets and then sampling the expectation values using quantum SSE.

We consider the computational cost of implementing quantum SSE for the special case (see Section~\ref{sec::quantumSSE}). In the quantum SSE algorithm outlined previously, the cost of directly sampling $\bra{\alpha_A}H_{b_n} \ldots H_{b_1} \ket{\alpha_A}$ given operator string $b$ requires $n$ unitary operations to be performed, multiplied by the number of samples $t$ for any target numerical precision.

Combining this with the fact that $\expval{n}$, the average expansion power in SSE, is proportional to the system energy and scales with $\beta N$, we see that the overall cost of sampling the configuration weight in the special case requires $\order{n} \sim \order{N} $ number of operations, i.e. it scales linearly with system size. A similar argument can also be made if we employ the amplitude estimation algorithm (see Appendix).

This is similar to the quantum SSE implementation for general Hamiltonians (see Section~\ref{sec::genHam}), where essentially the same set of unitary operations are performed, except with an additional control operation. We therefore expect the general implementation of quantum SSE to also scale with $\order{N}$.

We compare this to the classical version of the SSE algorithm. When there is no sign problem, the cost of sampling the configuration weight can be $\order{N}$. The classical algorithm in such cases can be highly efficient, and the quantum algorithm outlined above exhibits no obvious quantum advantage in terms of computational cost. In this case, the primary benefit of quantum SSE is that it allows more more observables $O$ to be measured compared to classical SSE implementations.

However, when the classical SSE method encounters the sign problem, the computational cost of avoiding negative probabilities is potentially $\sim e^{\order{N}}$. In comparison, the cost of implementing the quantum algorithm scales linearly with system size, so we expect the quantum advantage to be exponential.

\begin{figure*}[t]
	\centering
	\includegraphics[width=0.9\textwidth]{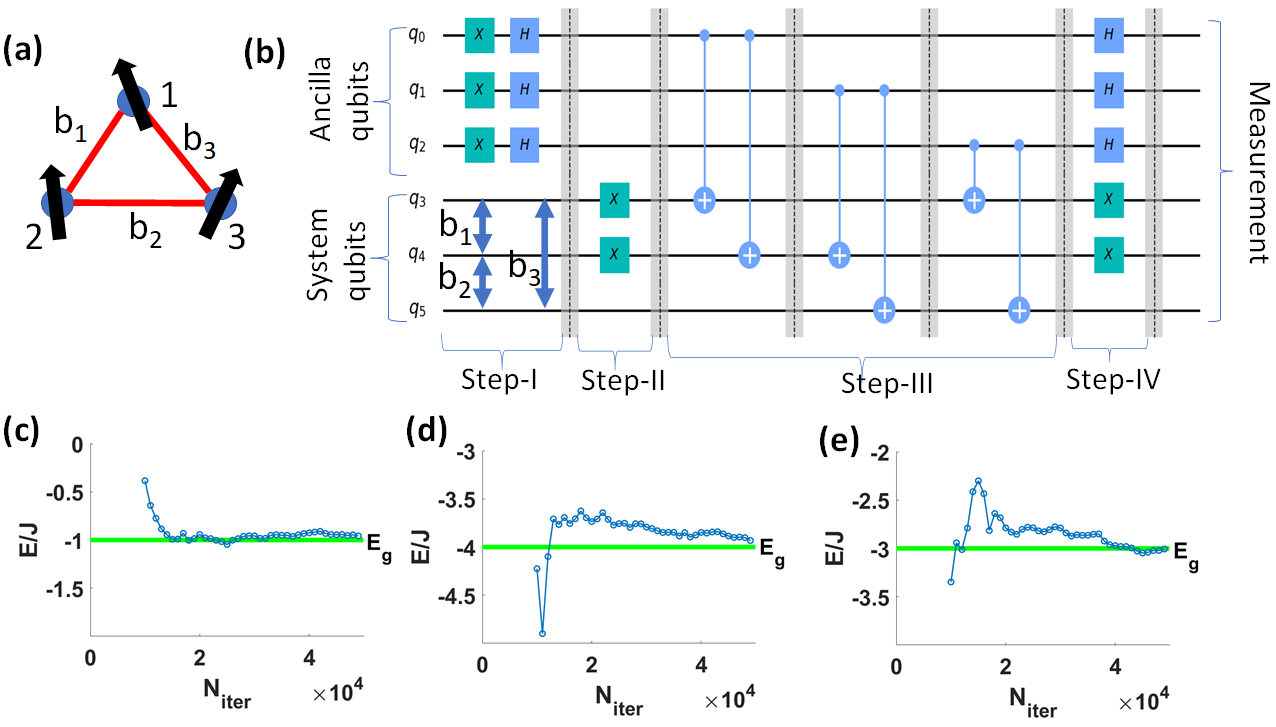}
	\caption{
		(a) 1D spin-1/2 chain with antiferromagnetic interaction and periodic boundary condition. For $N=3$, the sites are labelled $1,\,2,\,3$ and the corresponding bonds $b_1, b_2, b_3$.
		(b) An example schematic of the quantum circuit  calculating the expectation value of string of unitary operators $U^{b1}_{A,q1}U^{b2}_{A,q2}U^{b3}_{A,q3}$. Further details are given in the main text.
		(c), (d), (e) illustrates the convergence of the mean energy (blue-line with circles) determined by quantum SSE at $\beta=5$. The ground state energy represented by the green horizontal line is obtained via exact diagonalization. The $x$-axis indicates the number of Metropolis iterations $N_{\mathrm{iter}}$ for $N=3,\,4,\,5$ respectively.}
	\label{fig::numerical_result}
\end{figure*}

\section{Example}

As an example, we consider the Hamiltonian Eq.\ref{eq::XXInter} for one dimensional periodic spin chains with number of sites $N=3,\,4,\,5$. We then use the use the quantum simulation toolkit Qiskit\cite{Qiskit} to perform a numerical simulation of the algorithm and compare with exact results.
The Hamiltonian of the antiferromagnetic isotropic Heisenberg spin chain is given by
\begin{equation}
H'= J\sum_{b} \sigma_x^{b(1)} \sigma_x^{b(2)},
\end{equation}
where $J>0$ and $b(i)$ is the $i$-th site of the $b$-th bond (see Fig.~\ref{fig::numerical_result}a).
The classical SSE implementation violates the no-branching condition and may suffers from the sign problem when the basis states $\{ \ket{\alpha} \}$ are not product states pointing along the $x$-axis. In quantum SSE this is no longer a consideration as there is no longer a no-branching requirement and the string of bond operators have positive-semidefinite weights. To illustrate this, we choose our basis states $\ket{\alpha}$ to be product states pointing along the $z$-axis (i.e. products of $\ket{\uparrow},\ket{\downarrow}$).

After absorbing the negative sign in the Boltzmann factor (see Eq.\ref{eq::Boltzmann}) and adding identity operators to the bond operators to make the bond-operators positive-semidefinite, the effective Hamiltonian of the quantum SSE is (assuming $J=1$),
\begin{align}
H=\sum_b H_b,
\end{align}
where $H_b = \openone - \sigma_x^{b(1)} \sigma_x^{b(2)}$.

The unitary operator $U_{A,B_i}^b$ is defined by the map,
\begin{align}
U_{A,B_i}^b \ket{\alpha_A} \ket{-_{B_i}}
= \frac{1}{\sqrt{2}} \left ( \ket{\alpha_A} \ket{0_{B_i}} 
-\sigma_x^{b(1)} \sigma_x^{b(2)} \ket{\alpha_A} \ket{1_{B_i}} \right ).
\end{align}
The expectation value of a given string of bond operators $H_b$ is related to  $U_{A,B_i}^b$ via the expression:
\begin{align}
&\bra{\alpha_A} \bra{-_{B_1}} \ldots \bra{-_{B_n}} U_{A,B_n}^{b_{n'}}  \ldots U_{A,B_1}^{b_{1'}}  \ket{\alpha_{A}} \ket{-_{B_1}} \ldots \ket{-_{B_n}}\nonumber\\
&= \frac{1}{2^n} \bra{\alpha_A}H_{b_{n'}} \ldots H_{b_{1'}} \ket{\alpha_A}.
\label{eq::expectation2}
\end{align}

We illustrate the quantum circuit performing this measurement in Fig.~\ref{fig::numerical_result}b. The quantum circuit determines the expectation value of string of $U^{b3}_{A,q3}U^{b2}_{A,q2}U^{b1}_{A,q1}$ for a three site periodic system when $n=3$. We now describe in detail the steps involved in the circuit in Fig.\ref{fig::numerical_result}b. 

In Step I, the ancilla qubits $q_0$, $q_1$ and $q_3$ are prepared in the states $\ket{-_{q_0}},\,\ket{-_{q_1}},\,\ket{-_{q_2}}$ respectively, using Hadamard and Pauli X gates. 

In Step II the system qubits $q_3$, $q_4$ and $q_5$ representing the spin-1/2 sites of the physical spin chain are prepared in some product state (in this example the $\ket{\uparrow}$, $\ket{\uparrow}$ and $\ket{\downarrow}$ states) respectively using either the identity operation or the X-gate.

In Step III the unitary operators $U^{b3}_{A,q3}, \,U^{b2}_{A,q2}$ and $U^{b1}_{A,q1}$ are applied sequentially via CNOT operations.

Finally in Step IV, the qubits are rotated using Hadamard or X-gates and then measured in the computational basis. The probability of measuring all qubits with the outcome $0$ gives the square of the expectation value of  $U^{b3}_{A,q3}U^{b2}_{A,q2}U^{b1}_{A,q1}$.

After evaluating the expectation value for a given operator string and spin state, the weight factor $W(n,b,\alpha)$ can be determined using equation Eq.\ref{eq::weight}. The Metropolis algorithm, as described in section Sec.\ref{sec::metropolis}, is then implemented accordingly to update the quantum state and the operator string.
In SSE, the energy of the system can be efficiently evaluated using the expression\cite{Sandvik2010},
\begin{equation}
E=-\frac{\left\langle n\right\rangle}{\beta}+N, \label{eq::energySSE}
\end{equation}
where $\left\langle n\right\rangle$ is the average length of operator string per Metropolis loop. Note that the contributing term $N$ in Eq.~\ref{eq::energySSE} is due to adding a constant to the Hamiltonian to ensure positive semidefiniteness.

The energy calculations from quantum SSE as a function of the number of Metropolis iterations are shown Figs.\ref{fig::numerical_result}c,d,e for site numbers $N=3,\,4,\,5$ respectively at $\beta=5$.
We start the Metropolis sampling with some arbitrary string of operators and some arbitrary product state.
The average numbers operator string length $\left\langle n\right\rangle$ is then calculated after the initial $10^4$ Metropolis steps, and the mean energy is evaluated using Eq.~\ref{eq::energySSE}.
It can be seen that in all the cases considered, the mean energy computed via quantum SSE converges towards the exact ground state energy represented by the green line, which is obtained via exact diagonalization.

\section{Conclusion}

In this article, we proposed a possible quantum implementation of the SSE Monte Carlo algorithm and compare it to its classical counterpart. It is shown that in this case the cost of implementing a single Monte Carlo update scales linearly with the number of particles $N$. We compare this to the classical implementation of SSE, where certain many-body systems exhibit the sign problem. The existence of the sign problem incurs an additional cost that scales exponentially with $N$. The quantum algorithm avoids this by ensuring that the weight of the configuration is always positive, regardless of the chosen basis. This suggests that quantum computers can significantly speed up the simulation of complex quantum many body systems. Even when the sign problem is not present and classical SSE can be implemented efficiently, quantum SSE can still be advantageous, since it allows for more general observables to be measured. To illustrate this, we perform a numerical simulation of a 1D spin-1/2 chain using the quantum SSE algorithm in combination with a basis that is typically hard to implement using classical SSE methods. In all cases considered, it is shown that quantum SSE converges to the exact results obtained from exact diagonalization.

 It has been shown that a general solution the sign problem is in fact NP-complete \cite{Troyer2005}. The quantum SSE implementation discussed here can implement each Monte Carlo update in polynomial time, but that does not necessarily imply a polynomial time convergence of the statistical average $\expval{O}$ in general. We note that the NP hardness of the general sign problem is a statement about the convergence of statistical averages when configuration $C$ has negative weights. Here, we are instead comparing computational resources involved when performing a single Monte Carlo update in quantum versus classical SSE. Nonetheless, the quantum SSE algorithm shows that quantum computers are promising tools for accelerating the SSE Monte Carlo simulation in many scenarios. This may provide a pathway for probing the quantum properties of many body systems that are currently inaccessible to existing classical techniques.

\acknowledgments K.C. Tan was supported by the NTU Presidential Postdoctoral Fellowship program funded by Nanyang Technological University. Financial support from the Ministry of Education, Singapore, in the form of Grant No. MOE2018-T1-1-021 is gratefully acknowledged. We also acknowledge helpful discussions with A.W. Sandvik.

\appendix

\section{Amplitude estimation}

In the main text, we made use of a quantum subroutine called amplitude estimation, which is summarized by the following theorem.

\begin{theorem} [Amplitude estimation\cite{Brassard2002}]
Given one copy of a quantum state $\ket{\psi}$, and unitary transformations $U \coloneqq 2\ketbra{\psi}-\openone$ and $V = \openone - 2 P$, where $P$ is a projector satisfying $P^2 = P$, the amplitude estimation algorithm outputs an estimate $\hat{p}$ of $p = \ev{P}{\psi} $ such that 
\begin{align}
\abs{\hat{p}-p} \leq 2\pi \frac{\sqrt{p(1-p)}}{t} + \frac{\pi^2}{t^2}
\end{align} for any positive integer $t$ with success probability $1-\delta$ and $\delta \in (0,1)$. 

The amplitude estimation algorithm implements $U$ and $V$ a total of $mt$ times where $m$ is some multiplicative factor of order $\order{log\frac{1}{\delta}}$.
\end{theorem}

Let $\ket{\psi}= U_{A,B_n}  \ldots U_{A,B_1}  \ket{\alpha_{A}} \ket{+_{B_1}} \ldots \ket{+_{B_n}}$ and $P = \ketbra{\alpha}$ from the main text. This give us 
\begin{align}
p = \left [\frac{1}{2^n \abs{h_{b_n} \ldots h_{b_1}}} \bra{\alpha_A}H_{b_n} \ldots H_{b_1} \ket{\alpha_A} \right ]^2,
\end{align} so the algorithm actually outputs the square of required amplitude. However, since 
\begin{align}
\abs{\hat{p} - p} = \abs{(\sqrt{\hat{p}}-\sqrt{p})(\sqrt{\hat{p}}+\sqrt{p})} \geq  \abs{(\sqrt{\hat{p}}-\sqrt{p})}^2,
\end{align} we obtain the following bound for the amplitude
\begin{align}
\abs{\sqrt{\hat{p}}-\sqrt{p}}^2 \leq 2\pi \frac{\sqrt{p(1-p)}}{t} + \frac{\pi^2}{t^2}.
\end{align} 

Implementing the algorithms invokes unitaries $U$ and $V$ a total of $mt$ times each for any target precision and success probability. The overall complexity of the algorithm therefore depends on the complexity of performing $U$ and $V$. Now, let $\ket{\psi} = W \ket{0,\ldots ,0}$ for some unitary $W$. Then $U = W (2\ketbra{0,\ldots,0 } - \openone) W^\dagger$. To perform the unitary $(2\ketbra{0,\ldots,0 } - \openone)$ just requires you to check if every particle is in state $0$, which can be done using $\order{N+n}$ Toffoli gates. Since $n \sim \order{N}$. The cost of implementing $U$ then  boils down to the cost of performing $W$, which is the cost of preparing the state $\ket{\psi}= U_{A,B_n}  \ldots U_{A,B_1}  \ket{\alpha_{A}} \ket{+_{B_1}} \ldots \ket{+_{B_n}}$. This is also $\order{N}$ assuming the basis state $\ket{\alpha}$ can be efficiently prepared. An identical argument follows for $V$. The total cost of implementing the amplitude estimation algorithm therefore scales with $\order{N}$.

\end{document}